\documentclass[9pt,twocolumn,twoside]{osajnl}

\journal{ol} 

\setboolean{shortarticle}{true} 

\ifthenelse{\boolean{shortarticle}}{\colorlet{color2}{color2b}}{\colorlet{color2}{color2}} 

\title{Spectral selectivity in capillary dye lasers}
\usepackage{graphicx}
\author[1,2]{Esmaeil Mobini}
\author[1,2]{Behnam Abaie}
\author[1,2,*]{Arash Mafi}

\affil[1]{Department of Physics \& Astronomy, University of New Mexico, Albuquerque, NM 87131, USA}
\affil[2]{Center for High Technology Materials, University of New Mexico, Albuquerque, NM 87106, USA}

\affil[*]{Corresponding author: mafi@unm.edu}
\dates{Compiled \today}

\ociscodes{(140.2050) Dye lasers; (060.3510) Lasers fiber; (140.3410) Laser resonators.}

\doi{\url{http://dx.doi.org/10.1364/ao.XX.XXXXXX}}

\begin{abstract}
We explore the spectral properties of a capillary dye laser in the highly multimode regime. 
Our experiments indicate that the spectral behavior of the laser does not conform with a simple Fabry-Perot
analysis; rather, it is strongly dictated by a Vernier resonant mechanism involving multiple modes, which 
propagate with different group velocities. The laser operates over a very broad spectral range and the 
Vernier effect gives rise to a free spectral range which is orders of magnitude larger than that expected 
from a simple Fabry-Perot mechanism. The presented theoretical calculations confirm the experimental results. Propagating 
modes of the capillary fiber are calculated using the finite element method (FEM) and it is shown that the 
optical pathlengths resulting from simultaneous beatings of these modes are in close agreement with the optical 
pathlengths directly extracted from the Fourier Transform of the experimentally measured laser emission spectra.
\end{abstract}

\setboolean{displaycopyright}{true}

\begin{document}

\maketitle
\thispagestyle{fancy}

\ifthenelse{\boolean{shortarticle}}{\ifthenelse{\boolean{singlecolumn}}{\abscontentformatted}{\abscontent}}{}

Dye lasers have been a major player since mid 1960s with many attractive properties including a wide operating 
wavelength range, often spanning 50 to 100 nanometers, and can be reasonably efficient~\cite{Schafer,Bbpc}. 
However, because of the difficulties in handling the dyes which can be poisonous or even carcinogenic, and 
because of their rapid degradation during operation due to photo-bleaching, dye lasers have been mostly replaced with solid 
state lasers. Recent advances in optofluidic systems have brought attentions back to dye lasers 
again~\cite{Li2008}, primarily because dyes can be recirculated in such systems in enclosed set up mitigating
the disadvantages while benefiting directly from their most favorable properties.
Miniaturizing liquid dye lasers into a microfluidic device has many potential advantages
such as compactness, easy maintenance, safe laser operation, accurate spatial mode control, and low threshold energy~\cite{Li2008,helbo2003micro,whitesides2006origins,thorsen2002microfluidic,psaltis2006developing} with a wide range of applications 
including chemical and biological sensing~\cite{fan2011optofluidic,fan2014potential}. Microfluidic fiber lasers
have also appeared in optical fiber platform resulting in dye fiber lasers~\cite{Vasdekis:07,Gerosa:15,abaie2016observation,abaie2016random,lin2015manipulation,zhou2013fiber,sudirman2014all,he1995two}.
Fiber lasers based on capillary tubes and photonic crystal fibers filled with a dye solution were studied in detail by Vasdekis et al.~\cite{Vasdekis:07}. 
They reported a free spectral range (FSR) which was 300 times larger than what would be expected from a simple Fabry-Perot (FP) cavity analysis:
they attributed this spectral selectivity to a Vernier resonant mechanism between two transversely propagating modes in the waveguide.

In this paper, we explore the operation of a dye-filled fiber laser in the highly multi-modal regime, giving rise to multiple peaks in the laser 
emission spectrum, with a considerably larger spectral range ($\sim 50~{\rm nm}$) compared with the two-mode operation regime reported by Vasdekis et al.
We observe that a simple FP cavity analysis does not explain the FSR for the laser spectral modes, because it predicts spectral laser lines which are much more 
closely spaced than what we measure in the experiment. Here, we show that a Vernier resonant mechanism involving many operating modes is responsible for
the observed large value of the FSR. We develop a theoretical model which accurately predicts the FSR based on the geometrical and optical properties of 
the fiber and its propagating spatial modes. The calculated spectral features show a good quantitative agreement with the experiment, confirming the important 
role of the Vernier resonant mechanism in setting the spectral behavior of similar fiber-based systems.
\begin{figure*}[htp]
\centering
  \includegraphics[width=12 cm]{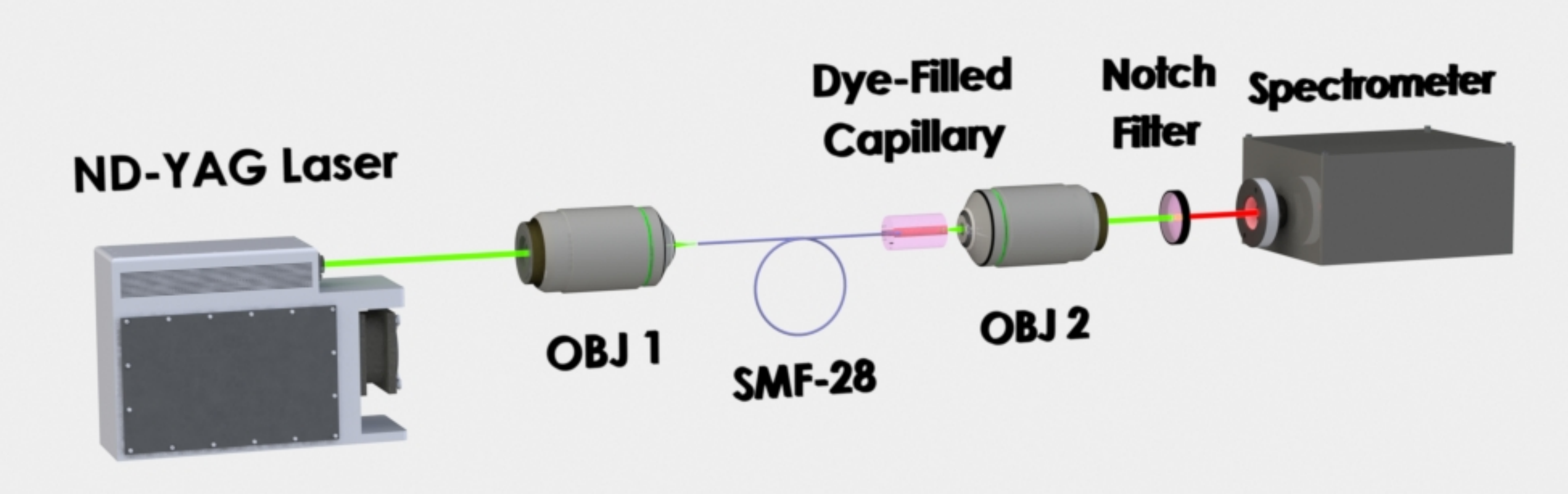}
\caption{A schematic of the experimental setup. Output of a pulse frequency doubled Nd:YAG laser is coupled into a single mode fiber. The single mode fiber is butt-coupled to the dye-filled capillary. The output is collected by a 40X microscope objective and sent into a spectrometer. Pump is filtered out using a notch filter.}
\label{fig:Set-up}
\end{figure*}

The fiber dye laser used here is a fused silica capillary of 5~\textmu m inner, and 300~\textmu m outer diameter, respectively. 
Rhodmaine 640 solution in benzyl alcohol at a concentration of 0.5~mg/ml was used to fill a few millimeter length of the fiber 
via capillary action. The benzyl alcohol solution and the cladding of the capillary have refractive indices of 
$n_{co}=1.53$ and $n_{cl}=1.46$, respectively. The relatively large refractive index contrast between fused silica cladding and 
benzyl solution in the fiber core enables the waveguide to operate in a highly multimodal regime. Calculations based on the $V$ number 
of the capillary waveguide indicate that there are nearly 70 transverse modes guided by the capillary waveguide~\cite{Saleh}.
The gain medium is pumped through an SMF-28 optical fiber to maximize the pump intensity overlap with the gain volume. The optical pump source 
is a 0.6~ns pulsed frequency-doubled Nd:YAG laser at a repetition rate of 50~Hz. The output is collected by a 40X microscope 
objective and is sent into a spectrometer as shown in Fig.~\ref{fig:Set-up}.

\begin{figure}[htbp]
  \centering
  \includegraphics[width=8.7cm]{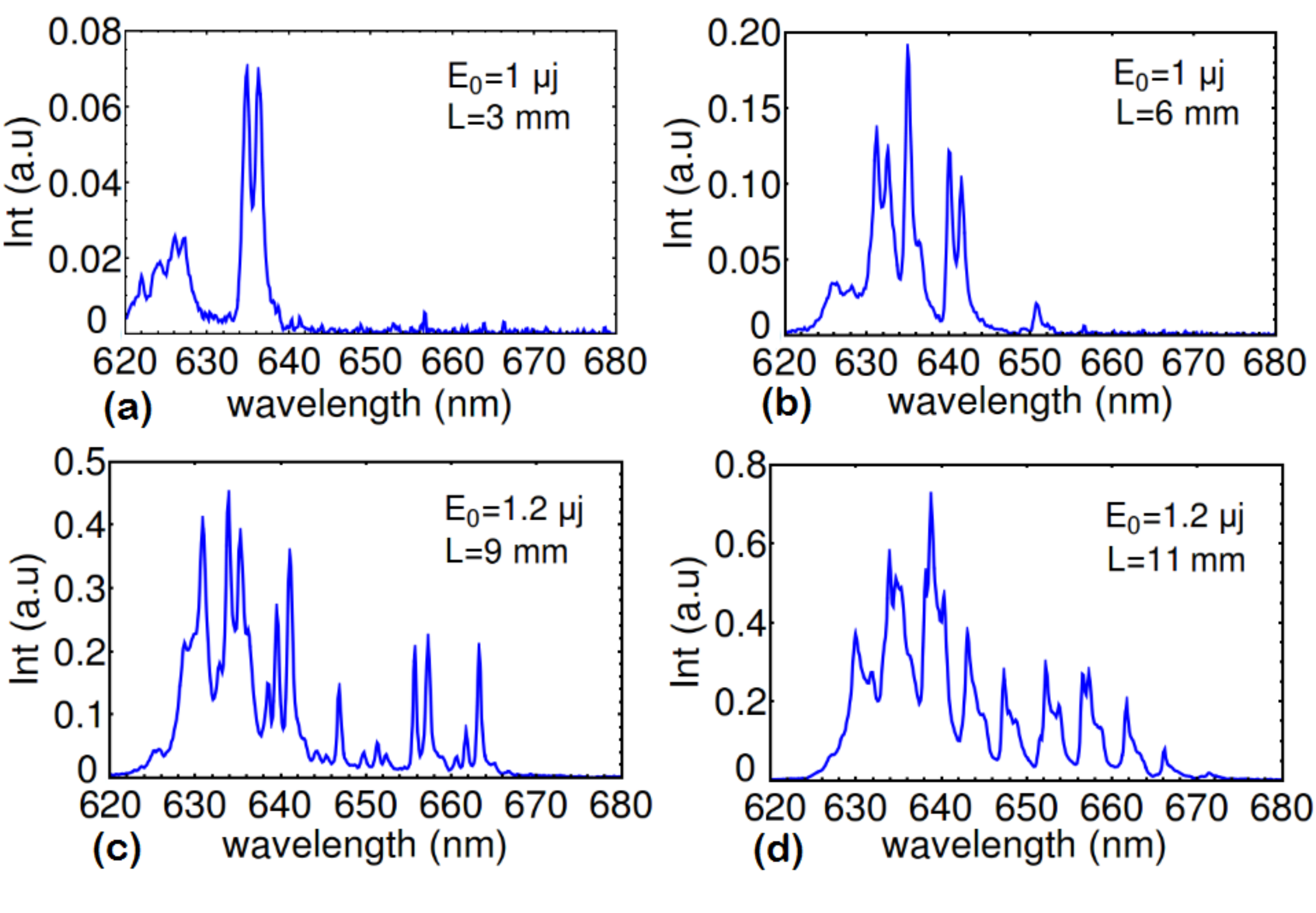}
\caption{(a)~Emission spectra of the capillary dye laser with an input pulse energy of $E_0$ and length of (a)~$L=3~{\rm mm}$, (b)~$L=6~{\rm mm}$, (c)~$L=9~{\rm mm},$ and (d)~$L=11~{\rm mm}$.}
\label{fig:Spect}
\end{figure}

Figures.~\ref{fig:Spect}(a)-(d) show the emission spectra of the capillary dye laser for four different lengths of the capillary.
Equation~\ref{eq:FSR} is commonly used to calculated the FP cavity FSR; using this equation for Fig.~\ref{fig:Spect}(d), e.g., gives
\begin{align}
\label{eq:FSR}
\Delta\lambda_{FSR}=\dfrac{\lambda^2}{2n_{co}L}\approx 0.014~{\rm nm},
\end{align}
which is clearly much smaller than the FSR obsereved in Fig.~\ref{fig:Spect}(d).

In order to investigate the spectral behavior of the capillary laser, we assume that there are several propagating modes undergoing multiple round 
trips inside the capillary. The end reflections are due to the air-fiber/solvent interfaces at the capillary tips. Ignoring the polarization dynamics and
assuming linearly polarized modes, we can expand the right-moving electric field propagating inside capillary as 
\begin{align}
\label{eq:fieldcom}
E(\rho,\varphi,z)=E_0\sum_{l=0}^{l_{max}}\sum_{m=1}^{m_{max}} A_{lm} J_{l}(\chi_{lm}\frac{\rho}{a})e^{il\varphi}e^{i\beta_{lm}z},
\end{align}
where $\beta_{lm}$ are the propagation constants of the modes. $J_{l}$ are Bessel functions of order $l$, which determine the transverse profile of the
linearly polarized $LP_{lm}$ modes. $\chi_{lm}$ is the m'th root of the $J_{l}$ Bessel function. $E_0$ is an overall constant, $A_{lm}$ is the complex amplitude coefficient
of each mode, $l$ ($l\in\{0,\sqrt{M}\}$) and $m$ ($m\in\{1,\sqrt{M}-l)/2\}$) describe the azimuthal and radial distributions 
of the electric field components, respectively. $M\approx {V^2}/{2}$ is the total number of guided modes, $V=k a \sqrt{n^2_{co}-n^2_{cl}}$ is the $V$ parameter of the waveguide, $k=2\pi/\lambda$, and $a$ is the radius of the capillary core~\cite{Okamoto,Saleh}.

The mode profiles and propagation constants can be numerically
solved for using analytical expressions.
In our analysis, we used a finite element method (FEM) code presented in Ref.~\cite{lenahan1983calculation}
to calculate the propagation constants of the modes to a high accuracy. 

Cavity loss from intrinsic attenuation and end mirrors are compensated by gain in the laser medium. 
Self-consistency requires the electric field of Eq.~\ref{eq:fieldcom} to be reproduced
after one round-trip. This translates into
a series of conditions expressed as
\begin{align}
\label{eq:selfcon1}
e^{i\beta_{lm}2L}=1,\quad {\rm for\ all\ \ } {l,m}, 
\end{align}
where $L$ is the total length of the capillary.
Of course, if a particular $A_{lm}$ is equal to zero, i.e. that spatial 
mode is not excited because of symmetry or other reasons, then the corresponding phase condition in Eq.~\ref{eq:selfcon1} does not 
need to be satisfied. Similarly, if the coefficient is small, i.e. the spatial mode carries a small power compared to other modes,
the corresponding condition is less relevant and can likely be ignored. Here, we assume that the dominant mode which is excited
in the capillary dye laser is the fundamental ${\rm LP}_{01}$ mode. Equation~\ref{eq:selfcon1} can be expressed alternatively as
\begin{subequations}
\begin{align}
\label{eq:selfcon21}
&e^{i\beta_{01}2L}=1,\\
\label{eq:selfcon22}
&e^{i(\beta_{01}-\beta_{lm})2L}=1,\quad {\rm for\ all\ \ } {l,m}. 
\end{align}
\end{subequations}
\begin{figure}[htbp]
  \centering
  \includegraphics[width=7.5 cm]{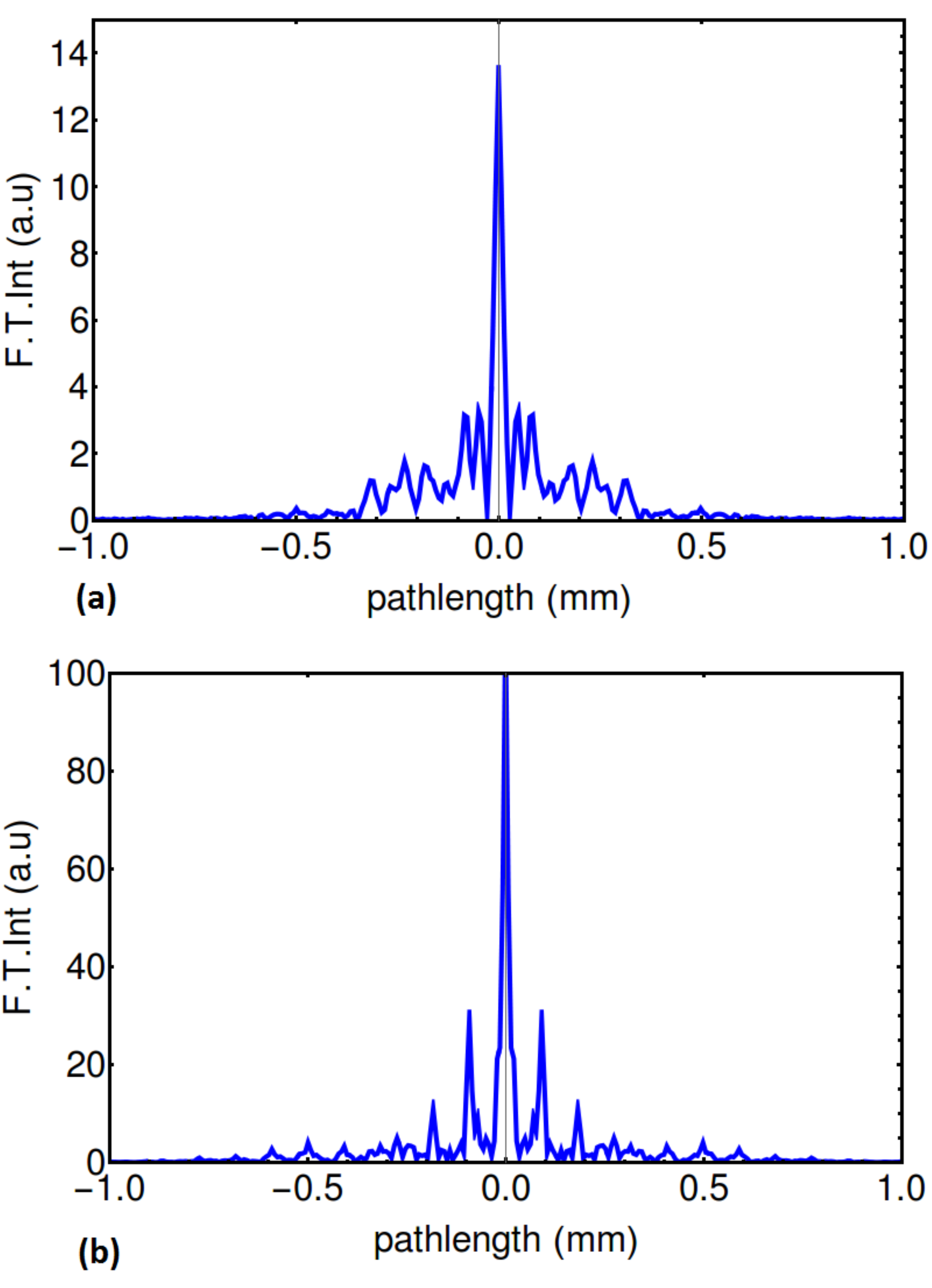}
\caption{(a)~Fourier transform of the emission spectrum of the 6~{\rm mm} capillary dye laser. (b) Fourier transform of the emission spectrum of the 11~{\rm mm} capillary dye laser.}
\label{fig:Ftspec}
\end{figure}
The explicit separation of the phase term corresponding to ${\rm LP}_{01}$ mode makes it more convenient to apply the assumption that
the ${\rm LP}_{01}$ mode is dominant in power. Equation~\ref{eq:selfcon21} results in a standard FP form expressed in Eq.~\ref{eq:FSR};
however, because the resulting FSR in our configuration is considerably below our spectrometer resolution, this term does not have any 
effect on the shape of the measured spectrum and is averaged out as observed experimentally. Rather, it is Eq.~\ref{eq:selfcon22} as the slow oscillating spectral term,
which controls the location of the spectral features in our experiment. 

In order to obtain the observed FSR from Eq.~\ref{eq:selfcon22}, let us consider the case where the phase matching is satisfied at frequency 
$\omega_q$, i.e. $[\beta_{01}(\omega_q)-\beta_{lm}(\omega_q)]2L=2q\pi$, and $q$ is an integer. The FSR is determined with $q+1$ at 
$\omega_{q+1}$, where a Taylor expansion results in
\begin{align}
\label{eq:FSRnew1}
\Big(\partial\beta_{01}/\partial\omega-\partial\beta_{lm}/\partial\omega\Big)\delta\omega=\pi/L.
\end{align}
This result can be cast in terms of the group indices of the modes
\begin{align}
\label{eq:FSRnew2}
(n^{(g)}_{01}-n^{(g)}_{lm})\delta\omega=\dfrac{\pi c}{L}\rightarrow\delta\omega_{lm}=\dfrac{\pi c}{\Delta n^{(g)}L},
\end{align}
where $\Delta n^{(g)}=n^{(g)}_{01}-n^{(g)}_{lm}$.

In order to quantitatively investigate the validity of Eq.~\ref{eq:FSRnew2}, we take the Fourier transform of the emission spectra of 
a 6~mm and an 11~mm capillary dye laser, as shown in Fig.~\ref{fig:Ftspec}~(a) and (b), respectively~\cite{Hofstetter,Polson,Vasdekis:07,camposeo2014random,poustie1994multiwavelength}. 
The Fourier transform is performed in the space of the wave-vector $k$; therefore, the Fourier conjugate space is characterized by the optical pathlength $d$,
whose peak positions $d_{lm}$ relate to the spectral selectivity in Eq.~\ref{eq:FSRnew2} by
\begin{equation}
\label{eq:optpath}
d_{lm}=2L\Delta n^{(g)}_{lm}=\dfrac{2\pi c}{\delta\omega_{lm}}.
\end{equation}
The peaks can be extracted from the Fourier transform data shown in Figs.~\ref{fig:Ftspec}(a) and (b).
A simple comparison between Figs.~\ref{fig:Ftspec}(a) and (b) shows that the spacing between two neighboring peaks 
resulted from the 11~mm capillary is nearly two times larger than that resulted from the 6~mm capillary. This linear 
relationship is a strong indication of the longitudinal nature of the spectral selectivity mechanism in the capillary laser~\cite{Vasdekis:07}.

A more concrete proof can be obtained by directly calculating the values of $d_{lm}$ using the fiber geometrical 
and optical parameters. Because we do not have access to the dispersion properties
of the dye material, we make a reasonable assumption that the values of the group- and phase- velocity 
differences are nearly identical. We use the values of
 $n_{co}$ and $n_{cl}$ reported earlier and the core radius to calculate the
propagation constants of the modes and establish the difference between the phase velocities of the fundamental mode and 
higher order modes. We note that by doing this, we are implicitly assuming that the majority of the laser power
is concentrated in the fundamental mode because of the strong central pumping. It is possible to look at other mode
combinations using the same formalism but we verified that our assumption captures the essence of the behavior
of this capillary laser to a high degree of accuracy.

\begin{figure}[htbp]
  \centering
  \includegraphics[width= 8 cm]{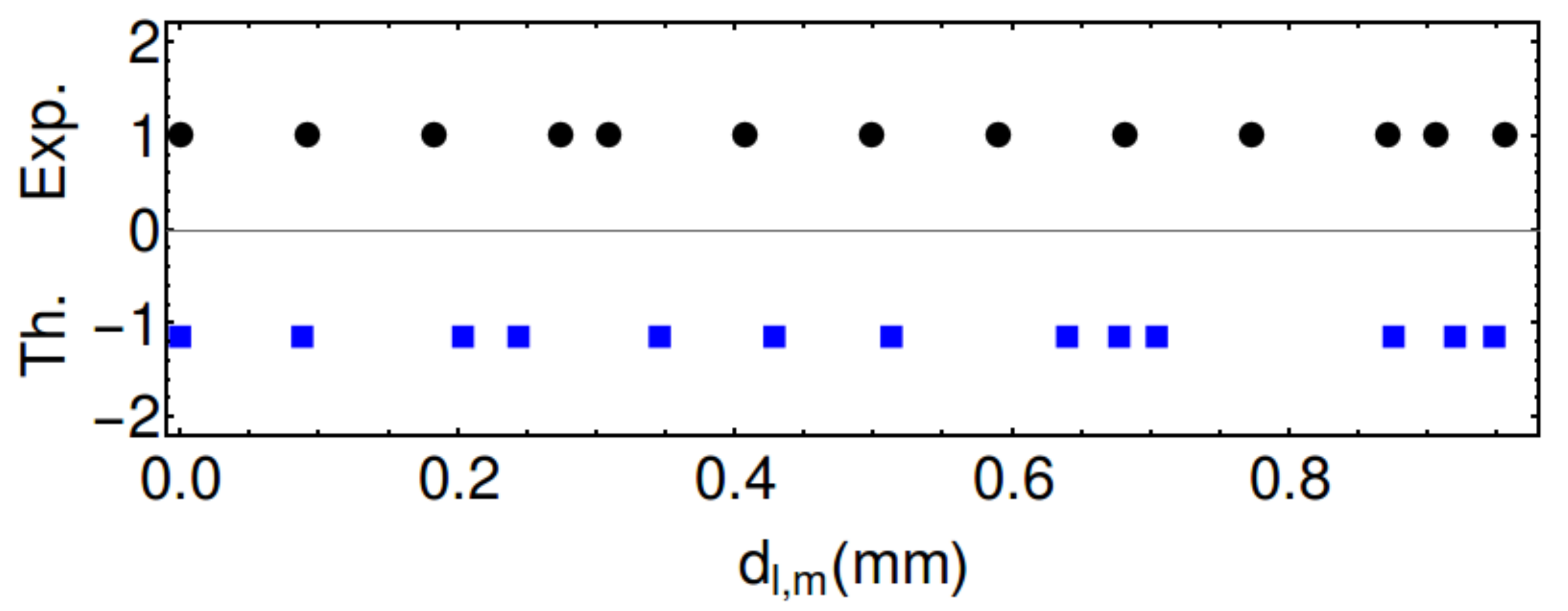}
\caption{The black column presents the extracted optical pathlengths from Fig.~\ref{fig:Ftspec} (b). The blue column presents 
the resulted optical pathlengths from the calculated propagation constants inside the capillary.}
\label{fig:Deln}
\end{figure}

In Fig.~\ref{fig:Deln}, we show the values of $d_{lm}$ directly calculated from the Fourier transform of the spectrum of 
the 11~mm capillary laser in Fig.~\ref{fig:Ftspec} (b): the values of $d_{lm}$ are extracted from the positions of the 
major peaks in the figure and shown as black circles in Fig.~\ref{fig:Deln}. Using the FEM code, we 
calculate $\Delta n^{(ph)}=n^{(ph)}_{01}-n^{(ph)}_{lm}$, which is the difference between the phase indices of 
the ${\rm LP}_{01}$ fundamental mode and the higher order modes. $\Delta n^{(ph)}$ is substituted in Eq.~\ref{eq:optpath} in place of
$\Delta n^{(g)}$, and the values of $d_{lm}$ are shown as blue squares in Fig.~\ref{fig:Deln}. The strong quantitative resemblance 
between the experimental results and theoretical simulations indicate that our model is correct. Another way to see this result 
is directly in the frequency space, where e.g. the near periodic position of the peaks in Fig.~\ref{fig:Spect} (d) with a
peak-to-peak separation of around $\Delta \lambda \approx 5~{\rm nm}$, is almost exactly as that calculated from the beating
between ${\rm LP}_{01}$ and ${\rm LP}_{11}$ modes. This agrees well with a phase refractive index difference of $\Delta n^{(ph)}_{11}\approx 0.004$ 
calculated using the FEM code. Here, we take the wavelength associated with the maximum peak in Fig.~\ref{fig:Spect} (d) which is around 640
 {\rm nm} to calculate the average FSR from Eq.~\ref{eq:FSRnew2}.  

Note that we only focused on the major peaks in Fig.~\ref{fig:Ftspec} (b)--minor peaks are likely related to the beating 
among modes, not including the ${\rm LP}_{01}$ fundamental mode. The major sources of discrepancy are likely: 1) the substitution 
of $\Delta n^{(ph)}$ for $\Delta n^{(g)}$, and 2) the beating among higher order modes which are ignored in this analysis.

In conclusion, we have reported a detailed analysis of a capillary dye laser with a highly multimodal regime of operation. 
The intensity spectrum of the laser shows a FSR, which is much larger than what is expected from a naive FP analysis. It is shown that
the spectral selectivity is dictated by a Vernier resonant mechanism, which is predominantly driven by the beating between the
${\rm LP}_{01}$ fundamental mode and higher order modes operating in the laser. The experimental values of the FSR have
a strong quantitative resemblance to those calculated using the theoretical predictions.

\end{document}